\documentclass{elsart}
\usepackage{epsf}
\usepackage{amsmath}
\usepackage{amssymb}
\usepackage{amsfonts}
\usepackage{graphicx}
\usepackage{graphics}
\usepackage{ulem}
\usepackage{epstopdf}
\usepackage{cite}

\newcommand{\xt}{x_{_\perp}}
\newcommand{\pt}{p_{_\perp}}
\newcommand{\ptjet}{p_{_\perp}^{~\rm jet}}
\newcommand{\ptphoton}{p_{_\perp}^{\gamma}}
\newcommand{\alphas}{\alpha_s}
\newcommand{\alphaem}{\alpha}
\newcommand{\sqrtsnn}{\sqrt{s_{_{\mathrm{NN}}}}}
\newcommand{\dd}{{\rm d}}
\newcommand{\X}{{\rm X}\,}
\newcommand{\A}{{\rm A}}

\newcommand{\qs}{Q_s}
\newcommand{\rag}{R^{^\A}_{_G}}
\newcommand{\rafd}{R^{^\A}_{_{F_2}}}
\newcommand{\ranmc}{R^{^{\rm NMC}}_{_{F_2}}}
\newcommand{\ga}{G^{^\A}}
\newcommand{\gp}{G^{^p}}
\newcommand{\fa}{F^{^\A}}
\newcommand{\fp}{F^{^p}}
\newcommand{\ria}{R_{_{i}}^{^\A}}
\newcommand{\rpa}{R_{_{p\A}}}
\newcommand{\rapprox}{R^{^{\rm approx}}}
\newcommand{\rppb}{R_{_{p {\rm Pb}}}}
\newcommand{\rdau}{R_{_{d {\rm Au}}}}
\newcommand{\gtarg}{G^{^{\rm targ}}}
\newcommand{\gproj}{G^{^{\rm proj}}}
\newcommand{\ftarg}{F^{^{\rm targ}}}
\newcommand{\fproj}{F^{^{\rm proj}}}
\def\cO#1{{{\cal{O}}}\left(#1\right)} 

\begin{document}

\vspace{-2.cm}
\begin{flushright}
\sffamily{CERN-PH-TH/2007-103, LAPTH-1202/07}
\end{flushright}

\begin{frontmatter}
\title{Measuring gluon shadowing\\[0.2cm]
  with prompt photons at RHIC and LHC}

\author[cern]{Fran\c{c}ois Arleo}\footnote{On leave from Laboratoire d'Annecy-le-Vieux de Physique Th\'eorique (LAPTH), Universit\'e de Savoie, CNRS, B.P. 110, 74941 Annecy-le-Vieux Cedex, France}\footnote{{\it Email address:} \texttt{arleo@cern.ch}}, 
\author[subatech]{Thierry Gousset}\footnote{{\it Email address:} \texttt{gousset@subatech.in2p3.fr}}

\address[cern]{CERN, PH Department, TH Division\\
1211 Geneva 23, Switzerland}

\address[subatech]{Subatech, UMR 6457, Universit\'e de Nantes,\\ Ecole des Mines de Nantes, IN2P3/CNRS.\\
 4 rue Alfred Kastler, 44307 Nantes cedex 3, France}

\begin{abstract}
The possibility to observe the nuclear modification of the gluon distribution at small-$x$ (gluon shadowing) using high-$\pt$ prompt photon production at RHIC and at LHC is discussed. The per-nucleon ratio, $\sigma(p+\A\to\gamma+X)/(A\times\sigma(p+p\to\gamma+\X))$, is computed for both inclusive and isolated prompt photons in perturbative QCD at NLO using different parametrizations of nuclear parton densities, in order to assess the visibility of the shadowing signal. The production of isolated photons turns out to be a promising channel which allows for a reliable extraction of the gluon density, $\rag$, and the structure function, $\rafd$, in a nucleus over that in a proton. Moreover, the production ratio of prompt photons at forward-over-backward rapidity in $p$--A collisions provides an estimate of $\rag$ (at small $x$) over $\rafd$ (at large $x$), without the need of $p$--$p$ reference data at the same energy.
\end{abstract}
\end{frontmatter}

\setcounter{footnote}{0}
\renewcommand{\thefootnote}{\arabic{footnote}} 	

\section{Why and how to probe gluon shadowing}

Gluon distributions in a proton and in a nucleus are fundamental ingredients in order to compute, within perturbative QCD (pQCD), hard-process observables in proton-proton, proton-nucleus, and nucleus-nucleus collisions. Thanks to years of detailed experimental studies at HERA and Tevatron supplemented by important theoretical developments in global fit analyses, the gluon distribution $\gp(x,Q^2)$ is fairly well known, say within a few percent accuracy, in the range $x\sim 10^{-5}$--$10^{-2}$ and $Q^2\sim 10$--$10^5$~GeV$^2$~\cite{Pumplin:2002vw} which is precisely the kinematical domain covered by most hard processes at the LHC. On the contrary, very little is known on the nuclear gluon density per nucleon, $\ga(x,Q^2)$~(see~\cite{Armesto:2006ph} for a recent review). So far, the only constraint on $\ga$ has been obtained from the scaling violation of the $F^{^\mathrm{Sn}}_{_2}/F^{^\mathrm{C}}_{_2}$ ratio~\cite{Gousset:1996xt} measured by NMC~\cite{Arneodo:1996ru}. The range of $x$ explored in this experiment is $0.02$--$0.2$, with a few GeV$^2$-wide $Q^2$-band around $Q^2=x\times 100$~GeV$^2$. However, the too large experimental uncertainties in these data do not allow for a precise estimate of the nuclear gluon distribution ratio,
\begin{equation}\label{eq:nuclear_distribution_ratio}
\rag(x,Q^2)=\ga(x,Q^2)\big/\gp(x,Q^2).
\end{equation}
The presently lack of knowledge on $\ga$ therefore prevents reliable pQCD predictions in high-energy nuclear reactions, such as $p$--Pb ($\sqrtsnn=8.8$~TeV) and Pb--Pb ($\sqrtsnn=5.5$~TeV) collisions at the LHC.

In addition to being a useful practical tool to predict hard processes in hadronic collisions, a precise knowledge of the gluon distribution is essential for a better understanding of evolution in QCD. In particular, an important theoretical activity has recently focused on possible non-linear QCD evolution at small $x$ and small $Q^2$, where the gluon density in the nucleus becomes large, $\ga(x,\qs^2)\sim 1/\alphas(\qs^2)$, and starts to saturate ($\qs$ stands for ``saturation scale'')~\cite{Iancu:2003xm}. Measuring $\ga$ in the vicinity of the saturation region, $Q\gtrsim \qs$, would therefore provide useful tests of the saturation picture.

For these reasons, it is essential to determine $\ga$ from the experimental data at RHIC and at the LHC with a similar accuracy to what has been achieved at HERA and Tevatron in the proton case. In $p$--A collisions, there are many ways to extract the nuclear gluon distribution $\ga(x,Q^2)$ or the nuclear gluon distribution ratio $\rag$. Let us discuss briefly various observables which look promising to achieve such a goal:
\begin{itemize}
\item The {\bf prompt photon} production channel\footnote{In the following, ``photons'' always refer to prompt photons. In particular, we do not consider photons coming from hadron decays.}\cite{Gupta:1987}, explored in detail in this Letter, has a rich phenomenology, ranging from fixed-target experiments to the Tevatron\footnote{At collider energies, the separation from $\pi^0$-decay photons becomes difficult without the use of isolation
criteria. This issue will be discussed further in Section~\ref{se:isolated}.}. Next-to-leading order (NLO) pQCD calculation provides an impressive description of the world-data on $\ptphoton> 5$~GeV hadroproduction spectra~\cite{Aurenche:2006vj}. As will be made more precise later, the mid-rapidity cross section is sensitive to parton distributions at  $x\sim \xt=2\,\pt/\sqrt{s}$ and $Q^2\sim \pt^2$. At a given $Q^2$, other values of $x$ can be probed by measuring photons at non-zero rapidity;
\item {\bf Jet} production is a direct competitor to the photon channel which has the clear advantage of large counting rates at the LHC. The lowest $\ptjet\gtrsim 30$~GeV above which the jet signal becomes tractable is, nevertheless, rather large as compared to that of photons ($\ptphoton\gtrsim 5$~GeV), making both observables complementary for the exploration of parton distributions. This is especially true for the typical $Q^2$-range probed by those two mechanisms.  However, since nuclear modifications of gluon densities prove more pronounced at low $Q^2$, the photon process appears to be more appropriate to determine gluon distribution ratios;
\item Large-$\pt$ {\bf dilepton} production at small invariant mass, $M\lesssim \pt$, has rather recently been put forward as a surrogate to photon production~\cite{Berger:1998ev,Fai:2004tg}. The main advantage of this channel is the absence of a large background. This channel is however suppressed by $\alphaem$ and $\alphaem^2$ as compared to photons and jets, respectively, leading to much smaller counting rates;
\item {\bf Heavy boson} ($Z^0$, $W^\pm$) production at the LHC has also been proposed to probe the shadowing of sea quarks (which are driven by gluons) at large scales $Q^2\sim 10^4$~GeV$^2$~\cite{Vogt:2000hp,Zhang:2002jf}, even though nuclear effects are not expected to be too pronounced in this region.
\end{itemize}

\begin{figure}[htb]
  \begin{center}
    \includegraphics[height=9.5cm]{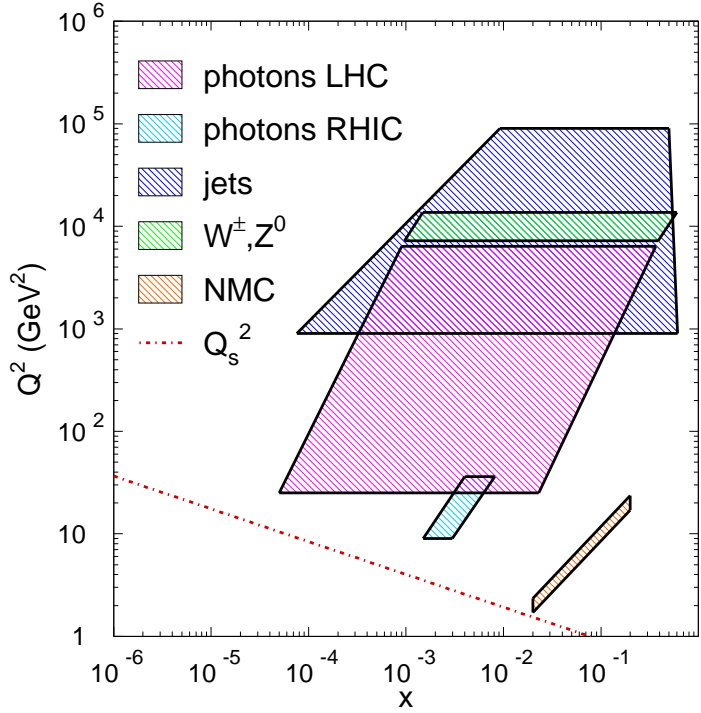}
  \end{center}
\caption{Typical $(x=\xt e^{-y}, Q^2=m_{_\perp}^2)$-domain probed in $p$-Pb collisions at RHIC and LHC using photon, jet, and heavy boson production. At LHC, photons produced at $\pt=5$--$100$~GeV and $|y|\le3$ are indicated, jets from $\pt=30$~GeV ($|y|\le4.5$) to $\pt=300$~GeV ($|y|\le2$), and heavy bosons at $\pt\le 100$~GeV and $|y|\le3$. At RHIC, photons produced at $\pt=3$--$6$~GeV and $y=2$--$3$ are selected. The kinematical range covered by the NMC experiment is also shown for comparison. The dash-dotted line indicates the saturation scale in a Pb-nucleus (see text for details).}
  \label{fig:xq2}
\end{figure}

The kinematical $(x, Q^2)$-domain probed with photons, jets, and heavy bosons (together with the range covered by NMC~\cite{Arneodo:1996ru}) is displayed in Fig.~\ref{fig:xq2}, which highlights the complementariness of these various probes. The dash-dotted line indicates the value of the saturation scale, $\qs^2=1\ {\rm GeV}^2\ (10^{-3}/x)^{0.3}$, obtained from the ``geometrical scaling'' fit to HERA data~\cite{Stasto:2000er}, scaled by $A^{1/3}=208^{1/3}$ in a Pb nucleus. As already pointed out, the photon channel covers in particular an important region at scales close to $\qs$ where shadowing is strongest.

The above list of observables is not meant to be comprehensive. In particular, more information could in principle be obtained from measurements of open heavy-flavoured mesons or large-$\pt$ light-hadrons. As we shall see later, however, the fragmentation process from partons to hadrons does not allow for the {\it partonic} kinematics to be determined. As a consequence, these processes rather probe moments of $\ga$ rather than $\ga$ itself.

So far, none of the above channels has produced any constraint on $\ga$ because of large error bars. At LHC, all these probes should be used to gain the most precise knowledge on parton distributions, and among them on $\ga$. First investigations fitting in such a program have been described in Ref.~\cite{Accardi:2004be}. Below, a deeper analysis of the photon production channel is outlined.\bigskip

\section{Inclusive photon production in $p$--A collisions}

\subsection{Perturbative dynamical processes}

At leading-order (LO), i.e.\ $\cO{\alpha \alphas}$, large-$\pt$ prompt photon production proceeds via $2\to{2}$ processes at the parton level. There are two channels in which the photon is produced {\it directly} in the partonic scattering:
\[
q(\bar q)+G\to \gamma+q(\bar q)\ \mathrm{(Compton)},\quad
q+\bar q\to  \gamma+G\ \mathrm{(annihilation)},
\]
yet the annihilation channel is less than a tenth of the Compton scattering below $\xt<0.1$, hence the interest of photon production for probing gluon densities. At this order, the cross section for the direct channel is an integral over both the projectile and target parton momentum fractions, $x_1$ and $x_2$, of the elementary cross section multiplied by the parton distribution functions (PDF), generically $f_i^{\rm proj}(x_1)$ for the projectile and $f_j^{\rm targ}(x_2)$ for the target, summed of the parton flavours $i$ and $j$. Because of the $2\to 2$ scattering kinematics, the cross section reduces to a single integral along the hyperbola
\[
\left(x_1-\frac{\xt}{2} e^y\right)
\left(x_2-\frac{\xt}{2} e^{-y}\right)=\left(\frac{\xt}{2}\right)^2
\]
in the $(x_1,x_2)$-plane. In order to understand the phenomenology discussed in the following, it is useful to have in mind some well-known basic trends deduced from the above equation. First, increasing $\xt$ means exploring larger $x_1,x_2$ (for instance the symmetric point, $x_1=x_2$ at $y=0$ reads $x_1=x_2=\xt$). Next, going
to positive $y$ corresponds to probe larger $x_1$ in the projectile (proton) and smaller $x_2$ in the target (nucleus).

In addition to the above direct process, a parton formed in a $2\to{2}$ process can fragment into a collinear photon,
\[
q_1+q_2\to q_3+q_4 \quad ; \quad  q_3 \to\gamma+\X,
\]
where $q$ stands here for a generic parton. As thoroughly discussed in Ref.~\cite{Aurenche:1998gv}, although the $2\to{2}$ scattering process is formally $\cO{\alphas^2}$, the parton-to-photon fragmentation mechanism $\cO{\alpha/\alphas}$ makes this channel to contribute also at leading-order. The cross section in this channel has the same ingredients as the direct process, supplemented by the fragmentation process. The latter is encoded in a fragmentation function $D_{q_3\to\gamma}(z)$, where the momentum-fraction $z$ of the parent-parton carried away by the photon has to be integrated from $\xt \cosh y$ to 1. Consequently, the production of a fragmentation photon at a given $\pt$ and $y$ probes parton distributions at larger momentum-fractions $(x_1/z, x_2/z)$ than at $(x_1, x_2)$ in the direct channel. This can be read off the hyperbola equation, after the change $\xt/2\to \xt/(2z)$ is implemented to take into account the effect of fragmentation in the kinematics.

Exploration of nuclear modifications to parton densities in such a framework simply consists in the replacement, say for the target, of the proton densities by the nuclear parton distribution functions (nPDF):
\[
f_i^p(x_2)\to f_i^\A(x_2).
\]
Since the cross section is actually known at NLO, $\cO{\alpha \alphas^2}$, the analysis will be carried out at this level of accuracy. 

\subsection{Nuclear production ratio at the LHC}

The inclusive photon --~that is summing direct and fragmentation channels~-- $\pt$ spectra has been computed in $p$--$p$ and $p$--A collisions at $8.8$~TeV~(LHC conditions) using the work of Ref.~\cite{Aurenche:1998gv}. In $p$--$p$ scattering, the NLO CTEQ6M parton densities have been used in the calculation~\cite{Pumplin:2002vw}, and the fragmentation functions of quarks and gluons into photons were taken from the NLO fit of $e^+e^-$ data carried out in Ref.~\cite{Bourhis:1997yu}. Regarding nuclear PDFs, $f^A$ is obtained from the average over the proton and neutron distributions in a nucleus with atomic mass $A$ and atomic number $Z$, $f_i^\A = Z\ f_i^{p/\A} + (A-Z)\ f_i^{n/\A}$. Several parametrizations of the ratio of the proton distribution in a nucleus, $f_i^{p/\A}$, over that of a ``free'' proton,
\begin{equation}\label{eq:pdfratios}
  \ria(x, Q^2) = f_i^{p/\A}(x, Q^2)\ \big/\ f_i^{p}(x, Q^2),
\end{equation}
are available for each parton flavour $i$: EKS~\cite{Eskola:1998df}, HKM~\cite{Hirai:2004wq}, nDS~\cite{deFlorian:2003qf}; yet only the latter group has performed a global data analysis at NLO accuracy, used in the present Letter\footnote{For completeness, we checked that the main conclusion of this study does not depend much on which parametrization is used in the calculation.}. An alternative fit, for which the depletion of gluon shadowing is somehow arbitrarily enhanced at small $x$ as compared to nDS, is proposed in~\cite{deFlorian:2003qf} and also considered for comparison (labelled nDSg in the following).

\begin{figure}[h]
    \begin{center}
      \includegraphics[height=6.4cm]{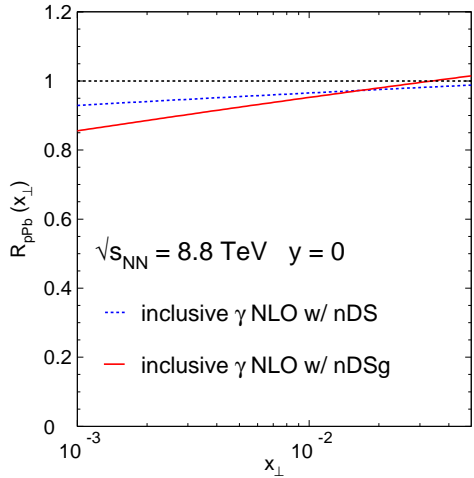}
    \end{center}
    \caption{Nuclear production ratio $\rppb$ of inclusive photon production at $y=0$ in $p$--Pb collisions at $\sqrtsnn=8.8$~TeV.}
    \label{fig:incy0lhc}
\end{figure}

In order to quantify nuclear effects, the nuclear production ratio
\begin{equation}
\label{eq:ratio}
  \rpa(\xt) = \frac{1}{A} \ \
  \frac{\dd^3\sigma}{\dd{y}\ \dd^2\pt}(p+\A\to\gamma+\X)
  \Big/ \frac{\dd^3\sigma}{\dd{y}\ \dd^2\pt}(p+p\to\gamma+\X)
\end{equation} 
is determined at LHC in the $\pt=5$--$100$~GeV range and at rapidity\footnote{In a $p$--Pb at $\sqrtsnn=8.8$~TeV at LHC, this corresponds to $y^{\rm lab}\approx 0.5$ in the laboratory frame. However, we checked that the results are practically unaffected by the small rapidity-shift from the centre-of-mass to the laboratory frame.} $y=0$. It is plotted in Fig.~\ref{fig:incy0lhc} as a function of $\xt$. $\rppb$ smoothly increases  from lower to higher $\xt$. At $\xt\simeq 10^{-3}$, the suppression is only roughly 10\% ($\rppb\simeq{0.92}$) when using nDS parton densities,  but proves larger ($\rppb\simeq{0.85}$) if the stronger nDSg gluon shadowing  is assumed in the calculation. At large $\xt$ (recall this would correspond  to large $\pt$ hence to large $Q^2$), there is basically no attenuation of the photon yield in $p$--Pb with respect to $p$--$p$ collisions. The trend of the inclusive photon suppression versus $\xt$ can be simply understood: as $\xt$ gets larger, the partons probed in the nucleus carry a higher momentum-fraction $x_2$ for which the gluon nuclear density ratio $\ria$ is less suppressed (i.e. closer to 1). Another reason comes from the $Q^2$-dependence of the shadowing process. Indeed, large-$\xt$ photons are produced perturbatively from highly-virtual partons $Q^2=\cO{\pt^2}$ (at fixed energy, $\xt$ and $Q^2$ are correlated) that are less affected by nuclear shadowing corrections, which die out at asymptotic $Q^2$. 

Similarly, this calculation is performed for inclusive photons produced at rapidity $y=2.5$. However, since the results are rather close to what is found in the isolated photon channel discussed later, photon production at forward rapidities is discussed in Section~\ref{se:isolated}.

Let us now briefly discuss to which extent such a suppression could be seen at the LHC. At mid-rapidity, the differential absolute cross section in $p$--Pb collisions is roughly
\begin{equation}
\begin{split}
  \frac{\dd\sigma^{p-{\rm Pb}}}{\dd{y}\ \dd\pt}&\Big|_{\pt=10\ {\rm GeV}} \simeq\ 1.4\ 10^7\ {\rm pb} / {\rm GeV},\\
  \frac{\dd\sigma^{p-{\rm Pb}}}{\dd{y}\ \dd\pt}&\Big|_{\pt=100\ {\rm GeV}} \simeq\ 0.9\ 10^3\ {\rm pb} / {\rm GeV}.
\end{split}
\end{equation}
This would correspond to a yearly rate of ${\cal N}\sim\ 10^7$ and $10^3$ per GeV-bin and per unit rapidity, at $\pt=10$ and $100$~GeV respectively, with the expected integrated luminosity in $p$--Pb collisions at the LHC ${\cal L}=1.4\ 10^{36}\ {\rm cm}^{-2}$ (taking 1 year~=~$10^6$~s)~\cite{Accardi:2004be}. Neglecting the statistical errors in the $p$-$p$ run taken at high luminosity, the statistical uncertainty on the nuclear production ratio is therefore negligible at $\pt=10$~GeV and roughly  $\delta \rppb\sim {\cal N}^{-1/2}\simeq 3\%$ in a 1~GeV-bin at $\pt=100$~GeV. Perhaps with the exception of the moderate suppression when using nDS for mid-rapidity photons, the fact that $\delta\rppb\ll 1-\rppb$ in all other cases indicates that the expected attenuation of inclusive photon at small $\xt$ could be observed experimentally, as far as statistical significance is concerned. Of course, this rough estimate does not include any consideration on detector acceptance\footnote{Note for instance the rather limited coverage of the PHOS electromagnetic calorimeter in the ALICE experiment, $\Delta y\times \Delta \phi = 0.26\times 1.7$.} and efficiency, nor possible systematic errors, which should be studied in detail in the experimental analysis. Nevertheless, we find it encouraging that, despite the much smaller photon rate than that of jets, the cross sections at the LHC are large enough to measure nuclear effects in this channel up to $\pt\sim 10^2$~GeV ($\xt\sim 2\ 10^{-2}$). Let us also mention that a similar accuracy is expected for photons produced at forward rapidity ($y=2.5$), which could be measured e.g. by the CMS experiment~\cite{lhcforward}.

\section{Extracting gluon nuclear distributions}
\label{se:isolated}

\subsection{Isolated photons and the relationship with gluon distributions}
\label{se:relationship}

We now turn to the study of isolated photon cross section. In this case, there is a very simple relationship between $\rpa(\xt)$ and the parton distribution ratios, Eq.~(\ref{eq:pdfratios}), $\ria(x,Q^2)$.  This relationship stems from the slow variation of the latter ratios as compared to those of the parton distributions themselves, $f_i^\A(x, Q^2)$. It is approximate, but the deviation from the exact result may well be hidden by other sources of uncertainty, in which circumstances such a direct extration is most valuable. As a matter of fact, the approximation can be gauged \textit{a posteriori}: the weaker the $\xt$-dependence of the measured ratio, $\rpa(\xt)$, the better the approximation.

The LO cross section is a sum of convolutions of projectile and target
parton densities with partonic cross sections. Writing $x_1=x_\perp
e^y/(2v)$ the LO Compton channel cross section reads:
\begin{eqnarray}
\frac{\dd^3\sigma}{\dd{y}\ \dd^2\pt}
=\frac{\alpha\ \alphas}{(\xt s/2)^2}
\int_{\xt e^y/2}^{1-\xt e^{-y}/2}\hspace{-5pt}
 &\dd{v}& \fproj\left(\frac{\xt e^y}{2v}\right)
\gtarg\left(\frac{\xt e^{-y}}{2(1-v)}\right)\,\frac{1}{3}
\left(1-v+\frac{1}{1-v}\right)\nonumber\\\label{eq:compton}
&&+\gproj\left(\frac{\xt e^y}{2v}\right)
\ftarg\left(\frac{\xt e^{-y}}{2(1-v)}\right)\,\frac{1}{3}
\left(v+\frac{1}{v}\right),
\end{eqnarray}
where $F(x)=F_{_2}(x)/x$. For proton-nucleus interaction, the per-nucleon cross section has the same expression with the changes
\[
\ftarg(x)\to \fa(x)=\rafd(x) \fp(x),\quad
\gtarg(x)\to \ga(x)=\rag(x) \gp(x),
\]
where we recall that $\fa$ and $\ga$ are understood as per-nucleon distributions\footnote{There is also a change $v_{\mathrm{max}}\to 1-\xt e^{-y}/(2 A)$, which is of no practical importance.}.

The integrand in~(\ref{eq:compton}) is strongly suppressed at the end-points. This is a result of the competition between $\fproj$ and $\gtarg$ (as well as $\gproj$ and $\ftarg$) that are suppressed at large values of their respective arguments. Since the PDF ratios, $\rafd$ and $\rag$, have a slow variation as compared to the $F\times G$ product, it can be assumed to be constant and put out of the $v$-integral. The typical $x$ at which $R$ is probed is easy to estimate at small-$\xt$, and for $|y|$ not too large, where the integrand is driven by the small-$x$ dependence of $F$'s and $G$'s. Considering $F(x)\sim A x^{-a}$ and $G(x)\sim B x^{-b}$ at small $x$, quantities such as $v^a (1-v)^b$ show up for $F\times G$ products. If the difference between $a$ and $b$ is disregarded, $F\times G$ is maximal at $v=1/2$ ($a,b >0$). This is not the whole story, since the $v$-dependence beside that of $F\times G$ shifts the maximum of the integrand away from $v=1/2$. However, this is partly counterbalanced by the fact that $a<b$ (around $x=10^{-2}$, $a\approx 1.3$ and $b\approx 1.7$). In the following the simple replacement
\[
R\left(\frac{\xt e^y}{2v}\right)\to R(\xt e^y).
\]
will be studied. 

Making this replacement in Eq.~(\ref{eq:compton}) leads to a very simple relationship between $\rpa$ and nuclear distribution ratios: $\rpa$ is an average of $\rafd$ and $\rag$ weighted by $y$-dependent coefficients that come from the relative importance of the two terms in Eq.~(\ref{eq:compton}) evaluated for $p$--$p$. Rather than quoting the full result, three simple cases of interest will be discussed. At $y=0$, thanks to the symmetry between the two terms in Eq.~(\ref{eq:compton}) for $p$--$p$, the weights are $0.5$, hence
\begin{equation}
  \label{eq:rpa_y0}
\rpa(\xt,y=0)\simeq 0.5\ \rafd(\xt)+ 0.5\ \rag(\xt).
\end{equation}

At large $y\ge y_0$ (in practice $y_0 \simeq 2.5$), i.e. at somewhat larger $x_1$ (notice that the above reasoning at {\it small} $x$ is not adequate to properly describe this), the extinction of gluons leads to a suppression of the second term in Eq.~(\ref{eq:compton}), leading to
\begin{equation}
  \label{eq:rpa_y2.5}
\rpa(\xt,y\ge y_0) \simeq \rag(\xt e^{-y}).
\end{equation}
Conversely, at $y\le -y_0$,
\begin{equation}
  \label{eq:rpa_y-2.5}
\rpa(\xt,y\le -y_0) \simeq \rafd(\xt e^y).
\end{equation}

\subsection{Results at LHC and RHIC}
\label{sub:results}

In order to check the accuracy of the assumptions of the former Section, the production of isolated photons is computed in $p$--Pb collisions at $\sqrtsnn=8.8$~TeV in pQCD at NLO, using nDSg nuclear parton densities. The isolation criterion was chosen as follows: a photon is said to be isolated if the total transverse partonic energy inside a cone of radius $R=\sqrt{{\Delta{y}}^2+{\Delta{\phi}}^2}=0.4$ around the photon is less than $10\%$ of its transverse momentum: $E_{_{\perp \rm part}}(R=0.4)\ \le\ 0.1\ p_{_{\perp \gamma}}$.

\begin{figure}[h]
  \begin{minipage}[t]{6.4cm}
    \begin{center}
      \includegraphics[height=6.4cm]{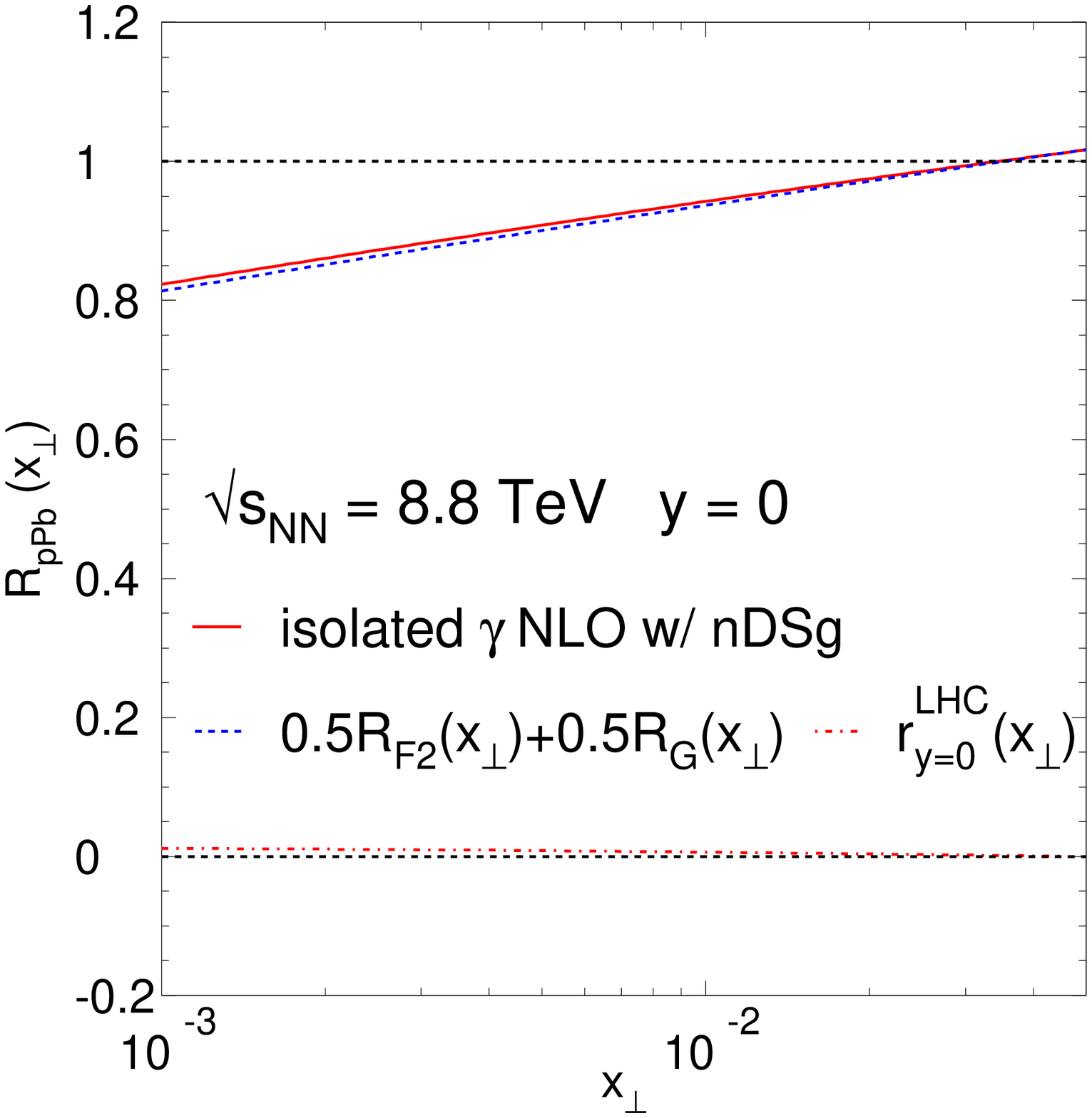}
    \end{center}
    \caption{Nuclear production ratio $\rppb$ of isolated photon production at $y=0$ in $p$--Pb collisions at $\sqrtsnn=8.8$~TeV.}
    \label{fig:isoy0lhc}
  \end{minipage}
  \hspace{0.4cm}
  \begin{minipage}[t]{6.4cm}
    \begin{center}
      \includegraphics[height=6.4cm]{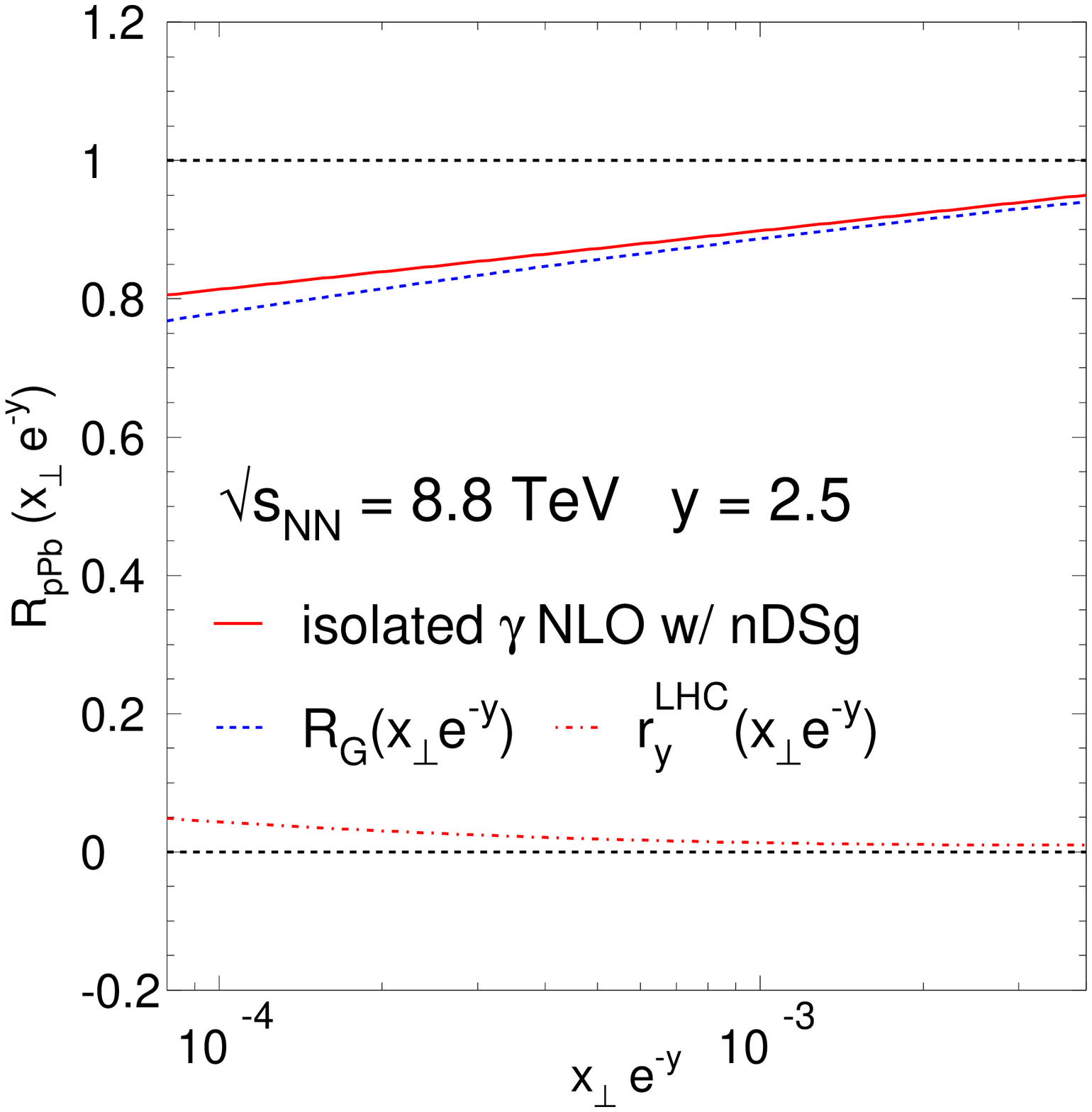}
    \end{center}
    \caption{Same as Fig.~\ref{fig:isoy0lhc} for photons with rapidity $y=2.5$.}
    \label{fig:isoy25lhc}
  \end{minipage}
\end{figure}

The nuclear production ratio at $y=0$ and $y=2.5$ is plotted as a function of $\xt e^{-y}$ in Fig.~\ref{fig:isoy0lhc} and Fig.~\ref{fig:isoy25lhc}. The $\xt$-dependence of $\rpa$ at $y=0$ is similar to what is observed in the inclusive channel, yet the suppression of the isolated photon yield, especially at low $\xt$, turns out to be slightly more pronounced. At mid-rapidity, the production ratio above $\xt\simeq 3\ 10^{-2}$ is consistent with 1, indicating small nuclear effects in the parton distributions.  The forward-rapidity photon suppression is larger, at a given $\xt$, than at mid-rapidity\footnote{If, on the contrary, one compares the suppression at a given $x^\prime=\xt e^{-y}$, the forward-rapidity photons are {\it less} suppressed than at $y=0$. Indeed, the larger the rapidity, the more virtual $Q^2=e^{2 y}\ {x^\prime}^2\ s/4$ the partons are probed in the nucleus, and therefore the lesser the nuclear shadowing.} and does not vanish even in the largest $\xt$-bin, $\xt e^{-y}\simeq 4\ 10^{-3}$. 

Let us now compare the isolated photon suppression, $\rpa$, with the analytic approximations $\rapprox_{y}$, Eqs.~(\ref{eq:rpa_y0}) to (\ref{eq:rpa_y-2.5}), in terms of parton densities ratios. In Fig.~\ref{fig:isoy0lhc}, $\rapprox_{y=0}(\xt)=0.5\ \left[\rag(\xt)+\rafd(\xt)\right]$ is plotted as a dotted line. The agreement between this approximation and the exact suppression ratio (solid line) is very good, of the order of one-percent accuracy on the full $\xt$ range, as can be seen from the ratio $r_{y=0}^{\rm LHC}=(\rpa-\rapprox_{y=0})/\rpa$ plotted as a dash-dotted line. Therefore, it turns out that neither the annihilation process nor next-to-leading order corrections, both neglected in Section~\ref{se:relationship}, spoil dramatically the quality of the analytic approximation. Consequently, the nuclear production ratio of isolated photons can serve as a reliable probe to measure small-$x$ shadowing of parton densities in large nuclei. The production of $y=2.5$ isolated photons shows a similar trend. At large rapidity, $\rpa$ can be approximated with $\rapprox_{y\gtrsim y_0}(\xt)=\rag(\xt e^{-y})$ plotted in Fig.~\ref{fig:isoy25lhc} (dotted line), together with $r_{y\gtrsim y_0}^{\rm LHC}=(\rpa-\rapprox_{y\gtrsim y_0})/\rpa$ (dash-dotted line), showing an agreement better than $5\%$. At both rapidities, a similar behaviour is observed: $\rpa$ is slightly above $\rapprox$ at small $x$ and then becomes closer to $\rapprox$ at large $\xt$. The fact that $\rpa\gtrsim\rapprox$ partly comes from the extra gluon radiated at NLO which requires the momentum-fraction carried by the target parton to be larger than it would at LO accuracy.

We now present the production ratio of isolated photons in $d$--Au (over $p$--$p$) collisions at RHIC top-energy, $\sqrtsnn=200$~GeV. The calculation has been carried out in the forward-rapidity region ($y=3$) in the $\pt=3$--$6$~GeV range, corresponding to $\xt e^{-y}=1.5\ 10^{-3}$--$3\  10^{-3}$. Note that measurements at such large rapidities, $y\simeq 3$, could be envisaged both with the PHENIX and the STAR detectors~\cite{rhicforward}.

\begin{figure}[h]
  \begin{minipage}[t]{6.4cm}
    \begin{center}
      \includegraphics[height=6.4cm]{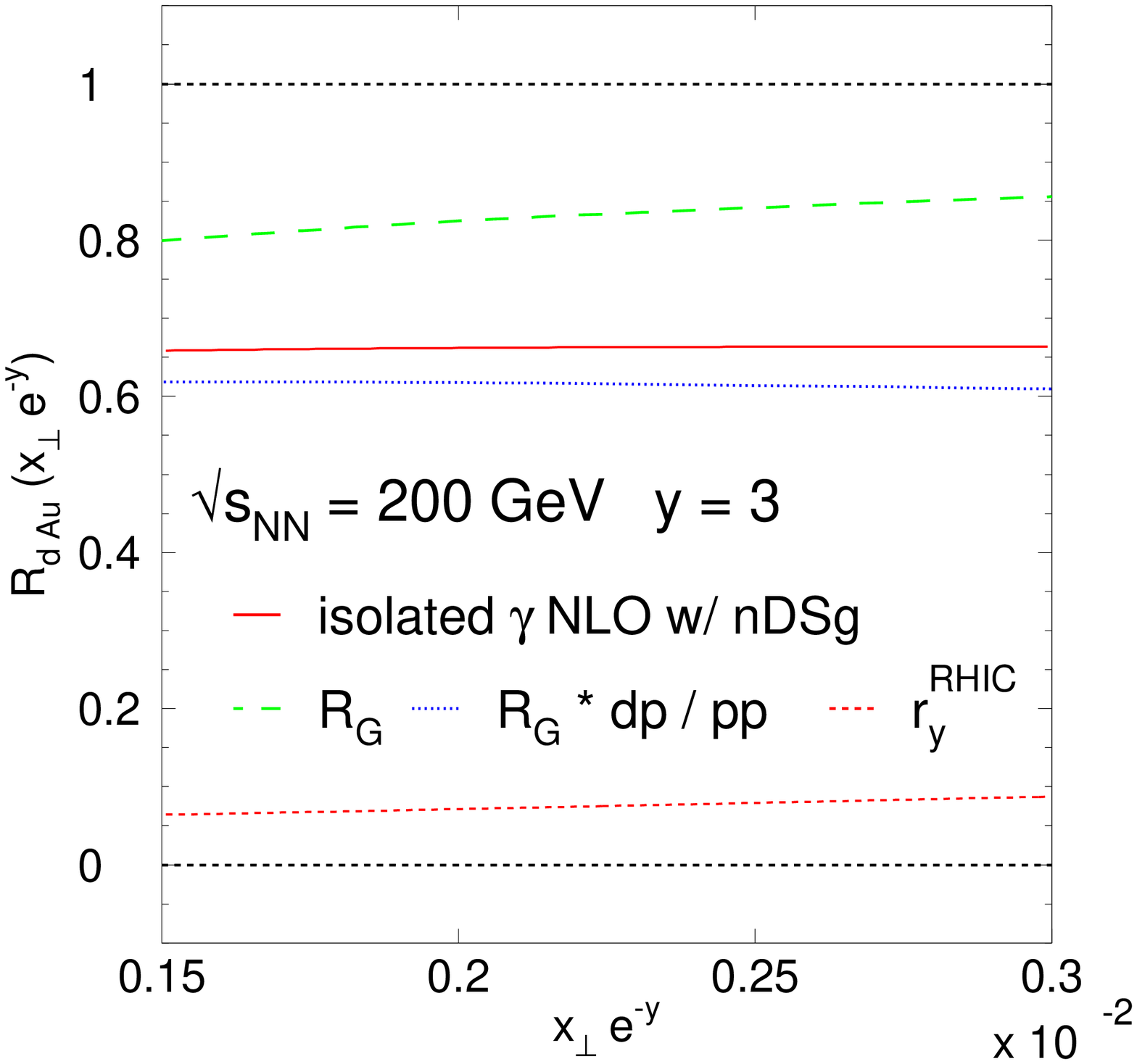}
    \end{center}
    \caption{Nuclear production ratio $\rdau$ of inclusive photon production at $y=3$ in $d$--Au collisions at $\sqrtsnn=200$~GeV.}
    \label{fig:isoy3rhic}
  \end{minipage}
  \hspace{0.4cm}
  \begin{minipage}[t]{6.4cm}
    \begin{center}
      \includegraphics[height=6.4cm]{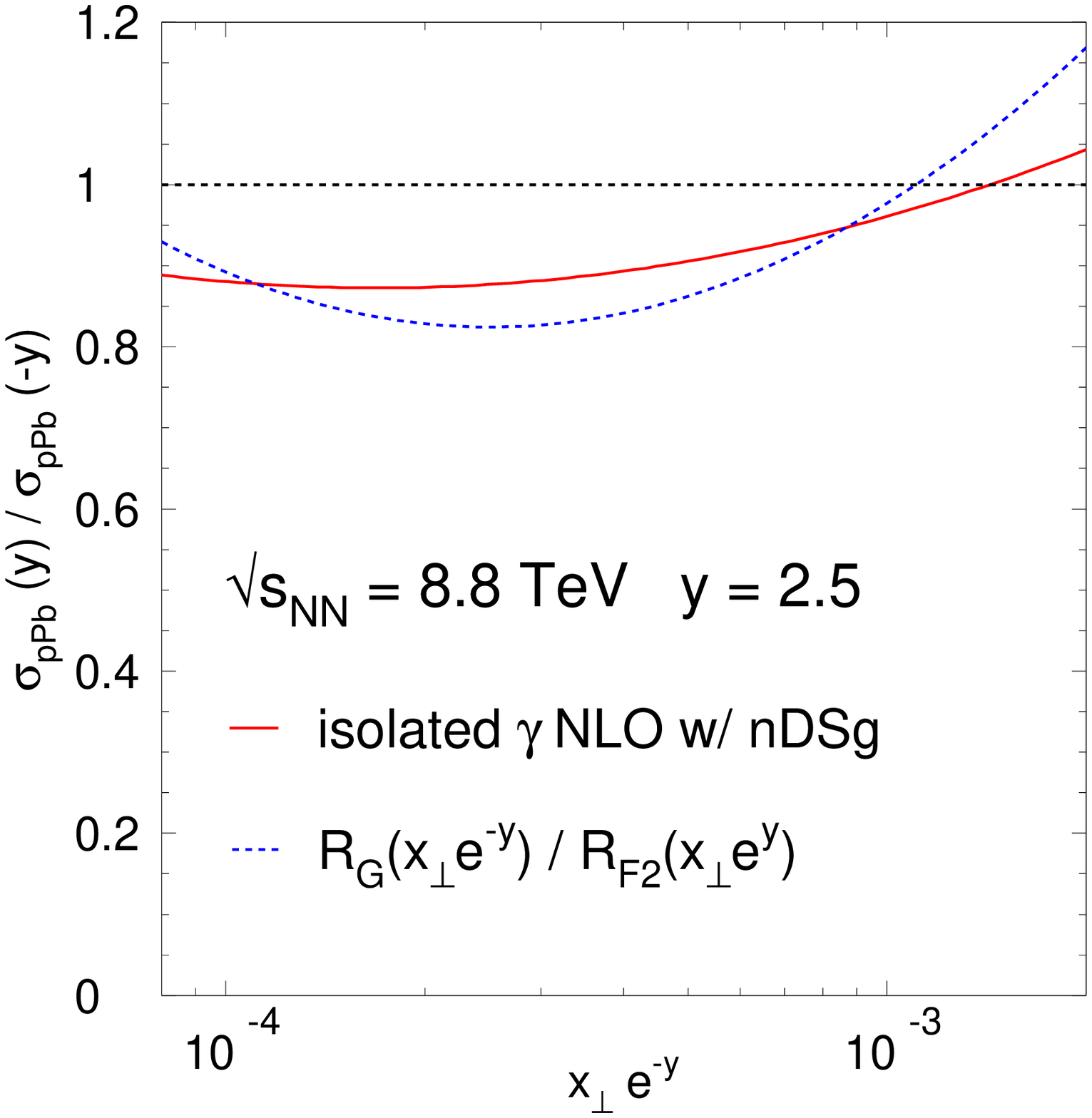}
    \end{center}
    \caption{Ratio of forward-over-backward $\sigma(+y)/\sigma(-y)$ inclusive photon production at $y=2.5$ in $p$--Pb collisions at $\sqrtsnn=8.8$~TeV.}
    \label{fig:fwdbwd}
  \end{minipage}
\end{figure}

Before we go on, let us stress that a reliable pQCD calculation can be seen as doubtful for two reasons. First of all, in that small $\pt$-domain (basically dictated by kinematic constraints), $\alpha_s$ is not too small and PDFs and FFs are not well constrained at such low virtualities. Moreover, the present fixed-order calculation might be somehow affected by large threshold and recoil resummation corrections at the edge of phase-space, that is at large $\pt$~\cite{Laenen:2000deLaenen:2000ij}. That said, let us explore how good the approximated expressions for $\rdau$ might be at RHIC within the current approach. The ratio $\rdau$, plotted in Fig.~\ref{fig:isoy3rhic} (solid line) together with the estimate $\rapprox_{y\gtrsim y_0}(\xt)=\rag(\xt e^{-y})$ (dashed), does not depend on $\xt$ in the limited coverage assumed in the calculation. The gluon distribution ratio, $\rag\simeq 0.8$, is similar to what is found at LHC and lie somewhat above $\rdau$. The disagreement between $\rdau$ and $\rapprox_{y\gtrsim y_0}$ is actually due to a trivial isospin effect in the deuteron projectile: the presence of the neutron, less efficient than a proton to produce prompt photons, leads to an additional suppression. This effect can however easily be corrected for, e.g. by multiplying $\rdau$ by the inclusive production ratio in $d$--$p$ over $p$--$p$ collision. This isospin-corrected ratio (labelled $R_{_G}*dp/pp$ and displayed as a dotted line) proves close to the measured $\rdau$ ratio, $r_{y\gtrsim y_0}^{\rm RHIC}=(\rdau-\rapprox_{y\gtrsim y_0}*dp/pp)/\rdau\simeq 8\%$, yet the agreement is somewhat less than at LHC energy.

\subsection{Extraction without a reference to $p$--$p$}

The direct connection between production ratios and nuclear distribution ratios is of course interesting. It shows the need for having the cross section in $p$--$p$ and $p$--A in the same kinematical conditions. At LHC, this necessitates an extrapolation in energy (the nominal center of mass energy is $14$~TeV in $p$--$p$, and $8.8$~TeV in $p$--Pb) and a comparison of event rates at different rapidities in the detector (in $p$--Pb the center of mass frame moves at a rapidity 0.47 in the detector frame). This gives extra sources of uncertainties in the extraction\footnote{In that sense, performing measurements in $d$--$d$ and $d$--Pb collisions (and to a lesser extent in $p$--$d$ and $p$--Pb) would have the double advantage to get rid of possible isospin effects when comparing nuclei with very different $Z/A$ ratios and to avoid the boost of the center-of-mass frame with respect to that of the laboratory.}.

There is a way out of this difficulty which can be obtained by comparing in the same experiment the $p$--A cross section at forward ($+y$) and backward ($-y$) rapidity. Thanks to the symmetry of the $p$--$p$ cross section in the change $y\to -y$, the ratio
\[
\frac{\sigma_{pA}(+y)}{\sigma_{pA}(-y)}
=\frac{\rpa(\xt,y)}{\rpa(\xt,-y)}
\approx \frac{\rag(\xt e^{-y})}{\rafd(\xt e^y)},
\]
the latter relationship coming from the approximations at $|y|\ge 2.5$ discussed previously. At large $y$, there is thus a way of obtaining $\rag$ at $\xt e^{-y}$ from the knowledge of $\rafd$ at $\xt e^{+y}$. As an illustration, the forward-over-backward ratio in $p$--Pb collisions is plotted in Fig.~\ref{fig:fwdbwd} together with the estimate $\rag(\xt e^{-y})/\rafd(\xt e^y)$. This approximation turns out to be good ($\sim 5\%$) as long as $\xt$ is not too large ($\xt e^{-y}<10^{-3}$).

At LHC, the $\pt$ range 5--100~GeV corresponds to $\xt e^{+y}\approx 0.01$--$0.03$ at $y=2.5$, that is precisely the NMC $x$-range (see Fig.~\ref{fig:xq2}). Making the $Q^2$-evolution of the NMC ratio to $Q^2=\pt^2$ (and in a larger nucleus) leads to the following estimate for $\rag$
\[
\rag(\xt e^{-y},\pt^2)\simeq\frac{\sigma_{pA}(\xt,y)}{\sigma_{pA}(\xt,-y)}\ \times\
\ranmc(\xt e^y,\pt^2)\quad\quad (y\ge 2.5)
\]

\section{Summary}
\label{sec:summary}

There is a major importance to determine precisely the nuclear gluon distributions at small $x$, whether it is to perform accurate pQCD predictions or to study QCD evolution equations. We explored in this Letter how prompt photon production in $p$--Pb collisions at LHC and $d$--Au collisions at RHIC can help to achieve this goal.

The nuclear production ratio of isolated photons in $p$--A collisions can be simply approximated as a linear combination of gluon distributions and structure functions in the nucleus A over those in a proton, $\rag$ and $\rafd$. Performing the calculation in pQCD at NLO we checked that such an approximation is correct up to a few-percent accuracy, making this observable an ideal tool to measure gluon shadowing at RHIC and at LHC. We also point out that the ratio $\rag(\xt\ e^{-y})/\rafd(\xt\ e^y)$ can be determined through the forward-over-backward photon production ratio in $p$--A collisions, without the need of any $p$--$p$ reference data at the same energy.

Although a particular set of the nuclear parton densities is used here as an example, the method and results obtained in this study are quite general. Present fits and models of nPDFs are not constrained in the small $(x, Q^2)$-domain, with a factor of 2 spread at $x\simeq 10^{-5}$~\cite{Armesto:2006ph}; the nDSg set being one of the fits with the weakest shadowing. Future prompt photon data, therefore with a suppression potentially larger than what is determined in this Letter, should be able to properly discriminate among the parameterizations currently available.

\providecommand{\href}[2]{#2}\begingroup\raggedright\endgroup

\end{document}